\documentclass[%
twocolumn,
nofootinbib,
 amsmath,amssymb,
aps,
]{revtex4-1}

\usepackage{graphicx}
\usepackage{dcolumn}
\usepackage{bm}
\usepackage[breaklinks=true,hidelinks]{hyperref}
\usepackage[mathlines]{lineno}
\usepackage[normalem]{ulem}

\usepackage[dvipsnames]{xcolor}

\begin{document}


\title{
	Multiscale simulation of the focused electron beam induced deposition process
}


\author{Pablo de Vera$^{1,2}$}
\email{pablo.vera@um.es}
\author{Martina Azzolini$^3$}
\author{Gennady Sushko$^1$}%
\author{Isabel Abril$^4$}
\author{Rafael Garcia-Molina$^2$}
\author{Maurizio Dapor$^3$}
\author{Ilia A. Solov'yov$^{5}$}
\email{On leave from A. F. Ioffe Physical-Technical Institute, 194021 St. Petersburg, Russia}
\author{Andrey V. Solov'yov$^{1}$}%
\email{On leave from A. F. Ioffe Physical-Technical Institute, 194021 St. Petersburg, Russia}
\affiliation{%
 $^1$MBN Research Center, Altenh\"{o}ferallee 3, 60438 Frankfurt am Main, Germany
}%
\affiliation{$^2$Departamento de F\'{i}sica -- Centro de Investigaci\'{o}n en \'{O}ptica y Nanof\'{i}sica (CIOyN), Universidad de Murcia, E-30100 Murcia, Spain
}%
\affiliation{$^3$European Centre for Theoretical Studies in Nuclear Physics and Related Areas (ECT*), I-38123 Trento, Italy
}%
\affiliation{$^4$Departament de F\'{i}sica Aplicada, Universitat d'Alacant, E-03080 Alacant, Spain
}%
\affiliation{%
 $^5$Department of Physics, Carl von Ossietzky University, Carl-von-Ossietzky Stra\ss e 9-11, 26129 Oldenburg, Germany
}%




\date{\today}

\begin{abstract}
	
	
Focused electron beam induced deposition (FEBID) is a powerful technique for 3D-printing of complex nanodevices.
However, for resolutions below 10 nm, it struggles to control size, morphology and composition of the structures, due to a lack of molecular-level understanding of the underlying irradiation-driven chemistry (IDC). Computational modelling is
a 
tool 
to 
comprehend and further optimise 
FEBID-related 
technologies.
Here we utilise a novel multiscale methodology which
couples Monte Carlo simulations for radiation transport with irradiation-driven molecular dynamics for simulating IDC with atomistic resolution.
Through an in depth analysis of W(CO)$_6$ deposition on SiO$_2$ and its subsequent irradiation with electrons, we provide a comprehensive description of the FEBID process and its intrinsic operation.
Our analysis reveals that 
these 
simulations 
deliver unprecedented results in modelling the FEBID process, 
demonstrating an excellent agreement with available experimental data of the simulated 
nanomaterial composition, microstructure and growth rate as a function of the primary beam parameters. 
\end{abstract}

\maketitle





Interaction of photon, neutron and charged particle beams with matter 
finds plenty of technological applications, 
particularly in materials science \cite{Robertson2012,Fowlkes2016}. Improvements in beam focusing and control are 
yielding cutting-edge methodologies for the fabrication of nanometre-size devices featuring unique electronic, magnetic, superconducting, mechanical and optical properties \cite{Sengupta2015,DeTeresa2016,Jesse2016,Fernandez-Pacheco2017,Winkler2017,Huth2018}. 
Among them, focused electron beam induced deposition (FEBID) is especially promising, as it enables 
reliable direct-write fabrication of complex, free-standing 3D nano-architectures 
\cite{Utke2008,Robertson2012,Huth2018}. Still, as the intended resolution falls below 10 nm, even 
FEBID struggles to 
yield
the desired size, shape and chemical composition \cite{Utke2008,Thorman2015,Shawrav2016}, which primarily originates from the lack of molecular-level understanding of the irradiation-driven chemistry (IDC) underlying 
nanostructure formation and growth \cite{Utke2008,Huth2012}. Further progress requires to learn how to finely control IDC.
This is achievable with the help of multiscale 
simulations \cite{DiazDeLaRubia2000,Sushko2016,Solovyov2017b}, provided that the model is sufficiently accurate and detailed, but also computationally feasible to allow exploring the wide range of deposition parameters.

FEBID operates through successive cycles of organometallic precursor molecules replenishment on a substrate and irradiation by a tightly-focused electron beam, which induces 
the release of organic ligands and the growth of metal-enriched nanodeposits.
It involves a complex interplay of 
phenomena, each of them requiring dedicated computational approaches:
(a) deposition, diffusion and desorption of precursor molecules on the substrate; 
(b) multiple scattering of the primary electrons (PE) through the substrate, 
with a fraction of them being reflected (backscattered electrons, BSE) and the generation of additional secondary electrons (SE) by ionisation; 
(c) electron-induced dissociation of the deposited molecules;  
and (d) the subsequent chemical reactions, 
along with potential thermo-mechanical effects \cite{Mutunga2019}. 
While processes (b) and (c) typically happen 
on the femtosecond-picosecond timescale, (a) and (d) 
may require up to 
microseconds or even longer.  Monte Carlo (MC) simulations have become an accurate tool for studying 
electron transport 
in condensed matter, 
and can also 
account for diffusion-reaction of molecules \cite{Smith2008,Plank2012,DaporBook}, but without offering atomistic details. 
At the atomic/molecular level, \textit{ab initio} methods permit the precise simulation of 
electronic transitions or chemical 
bond reorganisation \cite{Muthukumar2011,Stumpf2016,Rogers2018}, although their applicability is typically limited to the femtosecond--picosecond timescales and to relatively small molecular sizes. In between these approaches, classical molecular dynamics (MD) \cite{Solovyov2017b} and particularly reactive MD \cite{Sushko2015}
have proved to be very useful in the
atomistic-scale analysis of molecular fragmentation 
and chemical reactions 
up to nanoseconds 
and microseconds \cite{Sushko2015,deVera2019frag}. Still, a comprehensive and predictive multiscale simulation including all the FEBID-related processes has been, up to now, an elusive task.

A breakthrough into the atomistic description of FEBID was recently achieved \cite{Sushko2016} by means of the
new
 method that permitted simulations of irradiation-driven MD (IDMD) 
with the use of the software package MBN Explorer \cite{Solovyov2012}. 
IDMD superimposes probabilities of various quantum processes (e.g., ionisation, dissociative electron attachment) occurring in large and complex irradiated systems, stochastically introducing chemically reactive sites in the course of affordable reactive MD simulations. 
In the present investigation we utilise a combination of the aforementioned MC and IDMD methodologies and perform the first inclusive simulation of radiation transport and effects in a complex system where all the FEBID-related processes (deposition, irradiation, replenishment) are accounted for. Here specifically, 
detailed
space-energy distributions of electrons,  
obtained from MC \cite{DaporBook,Dapor2017,Azzolini2018} at different irradiation conditions, were used as an input for IDMD simulations \cite{Sushko2016,Solovyov2017b} on experimentally-relevant timescales,
where a direct comparison could be performed.

The coupled MC-IDMD approach was employed, for the first time, to analyse IDC at the atomistic level of detail for W(CO)$_6$ molecules deposited on hydroxylated SiO$_2$. In particular, the dependence on the primary beam energy and current of the surface morphology, composition and growth rate of the created nanostructures was analysed and was shown to be in an excellent agreement with results of available experiments \cite{Porrati2009}. This new methodology provides the necessary molecular-level insights into the key processes behind FEBID 
for its further 
development.
Furthermore, the approach being general and 
readily applicable to any combination of radiation type and material, opens unprecedented possibilities in the simulation of
many other problems where IDC 
plays an essential role, 
including astrochemistry 
\cite{Tielens2013,Mason2014}, nuclear and plasma physics 
\cite{DiazDeLaRubia2000}, 
radiotherapy \cite{Solovyov2017,Surdutovich2019} or photoelectrochemistry \cite{Zhang2016}.

\section*{Results and discussion}

Here we consider a multimolecular system,
consisting of 1--2 layers of W(CO)$_6$ molecules deposited on a 20$\times$20 nm$^2$ hydroxylated SiO$_2$ surface 
(in short, W(CO)$_6$@SiO$_2$), 
irradiated with PE beams of radius $R = 5$ nm
and energies $T_0=$ 0.5 -- 30 keV. 
This specific system is commonly used in FEBID and has been extensively studied experimentally \cite{Porrati2009,Fowlkes2010,Thorman2015} and theoretically \cite{Muthukumar2011,Sushko2016,deVera2019frag}. However, it has still been impossible to reach an adequate understanding of the process, such that to provide full control of the emerging nanostructures.

The electron transport in the substrate is treated by means of the MC program SEED \cite{Dapor2017,Azzolini2018}, which uses accurate inelastic \cite{deVera2013,deVera2015,deVera2019} and elastic \cite{Jablonski2004} cross sections for the interaction of electrons with condensed-phase materials as input parameters. 
Its coupling to MBN Explorer \cite{Solovyov2012} 
is done by providing energy- and space-dependent electron distributions, which determine the space-dependent rates for dissociation of molecules at the substrate surface. The interaction of the precursor molecules both with the substrate and with 
PE, BSE and SE is described by the IDMD method \cite{Sushko2016}. See the section ``Methods'' for further details.

In the next subsections, all
stages 
involved in the FEBID process of 
W(CO)$_6$@SiO$_2$
are individually studied and the parameters 
affecting the simulation of the whole process are determined.
Once this is done, a detailed analysis of the nanostructure growth rate, composition and microstructure as a function of the PE beam energy and current is performed.

\subsection*{\label{sec:diff}Precursor molecule interaction with the substrate}

The first factor affecting the nanostructure growth process 
is the ability of the precursor molecules to migrate to the irradiated area.
The surface diffusion coefficient
depends on the strength of the binding of the molecule to the surface, 
and could be determined experimentally \cite{Utke2008,Fowlkes2010}. 
However, 
this is not an easy task
for an arbitrary combination of precursor-substrate and temperature. 
Alternatively, molecular surface diffusion 
can be predicted by MD \cite{Sushko2016}
if the parameters for molecule-substrate interaction 
are known.
Here, we have simulated the diffusion of 
W(CO)$_6$@SiO$_2$ 
using the MBN Explorer software \cite{Solovyov2012} by means of
the procedure described earlier \cite{Sushko2016}. 
The obtained value of the diffusion coefficient at room temperature 
turned out to be 7.71 $\mu$m$^2$/s, 
being close to the experimentally determined value of 6.4 $ \mu$m$^2$/s \cite{Fowlkes2010}. 
See Supplementary Information S1 for further details.

\subsection*{\label{sec:MC}Electron beam interaction with the substrate}

The FEBID process is greatly influenced by the interaction between the PE beam and the substrate. PEs 
(of energies $T_0 = $ 0.5 -- 30 keV in the present investigation)
collide with precursor molecules, but also their multiple elastic and inelastic scattering in the substrate leads to the reflection of some of them (BSE), which re-emerge 
still keeping a significant fraction of their initial energy,
as well as to the ionisation of the medium and the production of a large number of SE with energies $T$ mainly in the 1--100 eV range. PE, BSE and SE
can interact with precursor molecules in very different ways, influencing the collision induced chemistry \cite{Thorman2015}, so it is essential to determine their yields and space and energy distributions.

\begin{figure}[t]
\includegraphics[
width=0.9\columnwidth
]{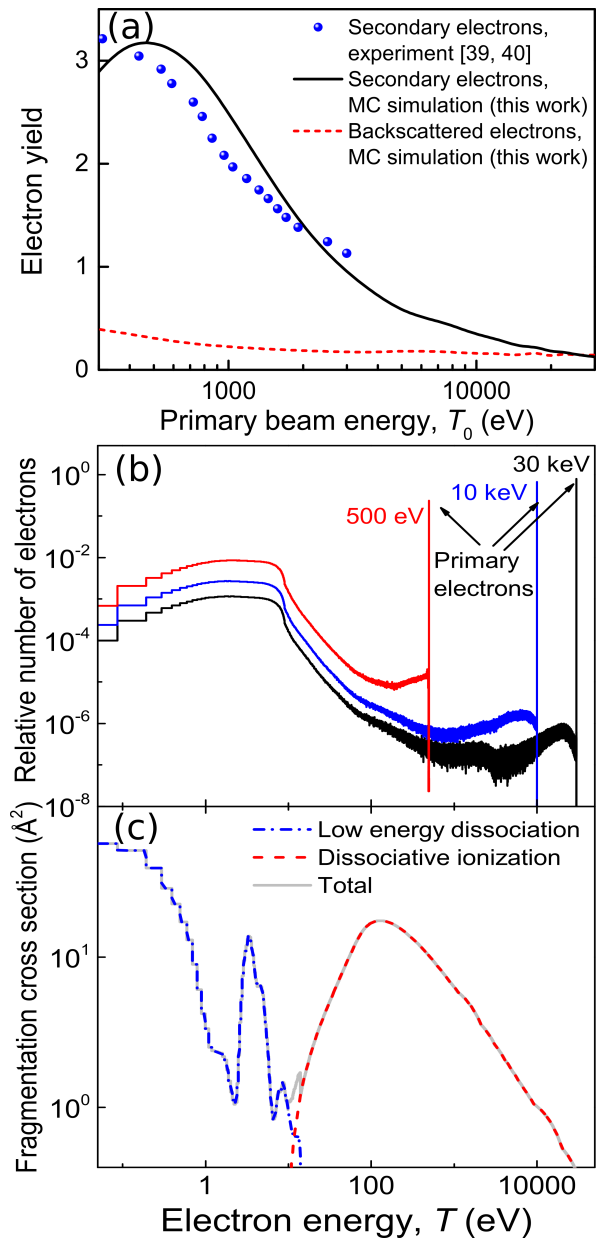}
\caption{\label{fig1:yields}
	(a) SE and BSE yields from SiO$_2$ as a function of PE energy. Symbols represent experimental data \cite{Glavatskikh2001,Yi2001}, while lines are the results from MC simulations.
The solid line shows the SE yield while the dashed line represents the BSE yield. 
(b) Energy distributions of SE and BSE crossing the SiO$_2$ surface for 500 eV, 10 keV and 30 keV PE. (c) Calculated electron-impact W(CO)$_6$ fragmentation cross section (solid line), with DI (dashed line) and low energy fragmentation (dash-dotted line) contributions. 
}
\end{figure}

MC simulations allow the analysis of the BSE and SE yields (total number of BSE and SE ejected per PE) as a function of the beam energy $T_0$. The SE yield is available experimentally for SiO$_2$ \cite{Glavatskikh2001,Yi2001} and is shown by symbols in Fig. \ref{fig1:yields}(a) 
together with the present simulation results (line),
which reproduce the main experimental features. 
The BSE yield is rather small, although comparable to the SE yield at large energies ($T_0\simeq 20-30$ keV).

Figure \ref{fig1:yields}(b) shows the relative number of electrons reaching the SiO$_2$ surface 
with different energies. It can be seen that, for all PE energies $T_0$, there is an intense SE peak at low energies, with its maximum at $T < 10$ eV, 
while the number of BSE (those with larger energies closer to $T_0$) is in general small. Further benchmarks of energy distributions against experimental data appear in Supplementary Information S2.

\begin{figure*}
\includegraphics[
width=2.0\columnwidth
]{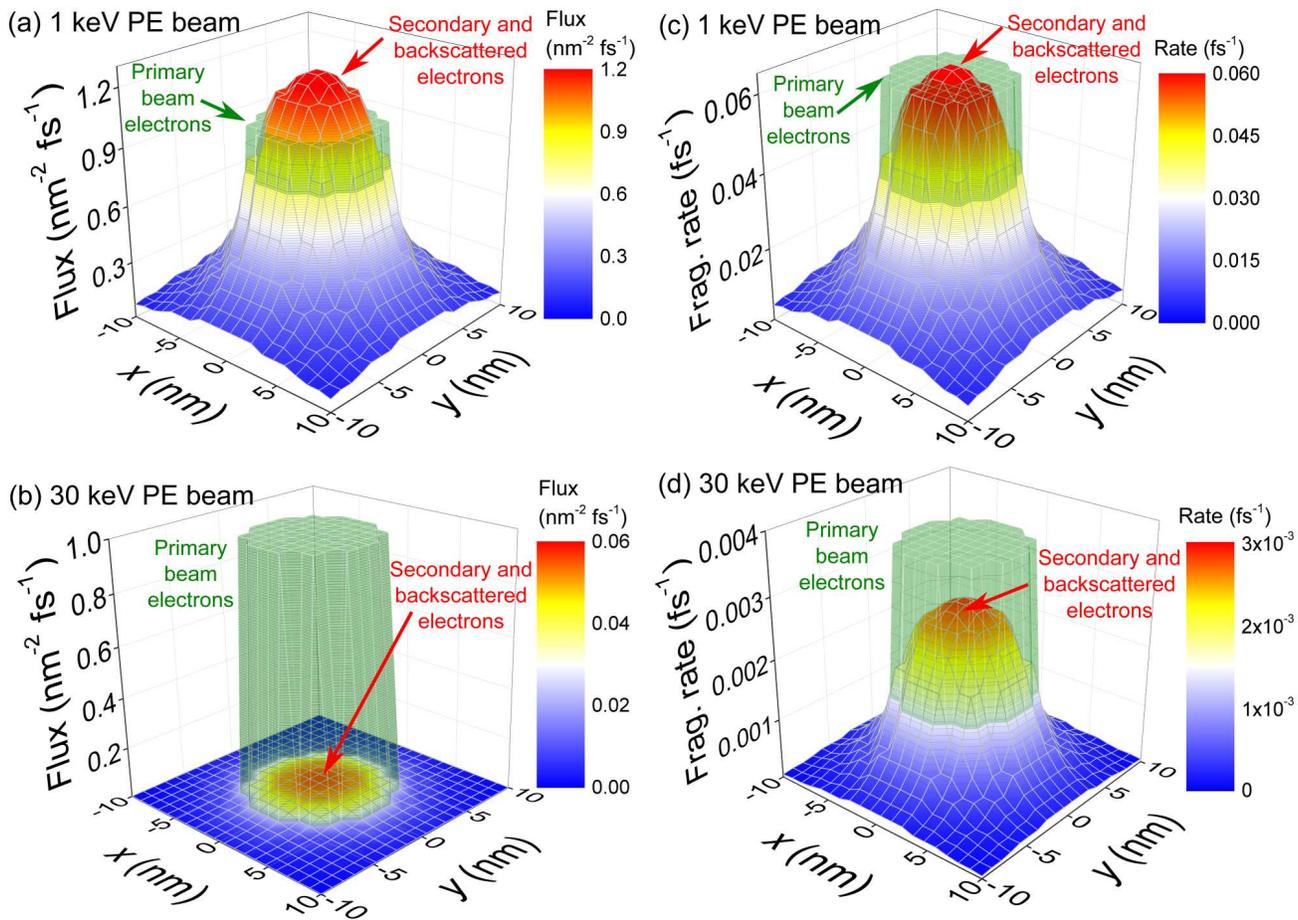}
\caption{\label{fig2:Nxy}
	(a)-(b) Space-dependent electron fluxes on a SiO$_2$ surface irradiated with a uniform PE beam of 5 nm radius, flux $J_0 = 1$ nm$^{-2}$ fs$^{-1}$ and energy of (a) 1 keV and (b) 30 keV. The green transparent surface depicts the PE flux in the beam area, while the coloured surface shows that due to SE and BSE. (c)-(d) W(CO)$_6$ fragmentation rate for (c) 1 keV and (d) 30 keV PE beams. 
}
\end{figure*}

MC simulations also provide the space- and energy-dependent fluxes $J(x,y,T)$ (electrons per unit area and unit time) of BSE and SE crossing the SiO$_2$ surface 
at different positions. 
These are shown in Figs. \ref{fig2:Nxy}(a) and (b) 
for uniform PE beams of 1 keV and 30 keV, respectively, and unit PE fluxes $J_0 = 1$ nm$^{-2}$fs$^{-1}$ within a circular area of radius $R=5$ nm.
While the high energy 30 keV beam produces a small number of SE and BSE everywhere, the lower energy 1 keV beam produces a large number of SE and BSE, which spread
outside the area covered by the PE beam
and exceed the number of PE at the centre of the beam.  

\subsection*{\label{sec:XS}Electron-impact molecular fragmentation cross sections}

Not only the number of electrons influences the properties of the structures emerging on the surface, but also the energy-dependent probability for W(CO)$_6$ molecule fragmentation, given by the corresponding cross section $\sigma_{\rm frag}(T)$, has an impact. This cross section includes dissociative ionisation (DI) 
for energies above the ionisation threshold ($\sim 8.5$ eV \cite{Wnorowski2012a}) 
as well as dissociative electronic excitations and dissociative electron attachment  
\cite{Thorman2015}.

Measurement of 
$\sigma_{\rm frag}(T)$
for the molecular fragmentation channels 
on the substrate 
is rather complicated, since the influence of all PE, BSE and SE crossing the surface cannot be disentangled. Under these conditions, what is usually measured is an effective decomposition cross section due to a PE beam of energy $T_0$, $\sigma_{\rm decomp}(T_0)$. 
Alternatively, gas-phase data may be used as a first approximation for the actual cross section $\sigma_{\rm frag}(T)$. 
For W(CO)$_6$ molecules, experimental information is available for DI \cite{Wnorowski2012a} and lower energy dissociation channels \cite{Wnorowski2012} \textit{relative} cross sections, 
but not the \textit{absolute} values needed for our simulations. 
The absolute DI cross section can be calculated by means of the dielectric formalism \cite{deVera2019}. 
The corresponding result is shown in Fig. \ref{fig1:yields}(c) by a dashed line.
For energies below 14 eV,
the experimental relative cross sections  \cite{Wnorowski2012} can be scaled in order to get a decomposition cross section $\sigma_{\rm decomp}(T_0)$
for 30 keV electrons incident in 
W(CO)$_6$@SiO$_2$
coinciding with the experimentally reported value 
\cite{Hoyle1994} (see Supplementary Information S3 for the details of the scaling procedure).   
The resulting 
low energy and total fragmentation cross sections appear in Fig. \ref{fig1:yields}(c) as dash-dotted and solid lines, respectively. 
DI dominates above $\sim$ 12 eV, while a large fraction of SE will fragment precursor molecules through the lower energy dissociation channels. 

\subsection*{\label{sec:FEBID}Simulation of the FEBID process}

The FEBID process relies on successive cycles of electron irradiation and precursor molecule replenishment \cite{Utke2008,Huth2018}. The irradiation phases are simulated by means of the IDMD method \cite{Sushko2016} by evaluating space-dependent bond dissociation rates for molecules on the substrate, which are calculated as explained in section ``Methods''. In brief, these
rates 
depend, in steady-state conditions, on both (i) the number and energies of the electrons crossing the SiO$_2$ surface at each point per unit time and unit area (which in turn are determined by the PE beam energy $T_0$ and flux $J_0$, see section ``Electron beam interaction with the substrate''), and (ii) the energy-dependent molecular fragmentation cross section $\sigma_{\rm frag}(T)$ (determined in section ``Electron-impact molecular fragmentation cross section'').

Figures \ref{fig2:Nxy}(c) and (d) illustrate the space-dependent fragmentation rates induced by uniform 1 keV and 30 keV beams, respectively,
of unit PE flux $J_0 = 1$ nm$^{-2}$fs$^{-1}$ within a circular area of radius $R=5$ nm. Although the number of BSE/SE electrons for 30 keV is small, their large cross section (in relation to PE) produces a significant fragmentation probability, but less than that due to PE at the beam area. However, for 1 keV, the fragmentation probability due to BSE/SE ($\sim$ 80--90 \% exclusively due to SE) is very large, and significantly extends beyond the PE beam area. These results clearly demonstrate the very different scenarios to be expected for beams of different energies and which will importantly influence the deposit properties, as well as the prominent role of low-energy SE on molecular fragmentation.

\begin{figure}[t]
	\includegraphics[
	width=0.9\columnwidth
	]{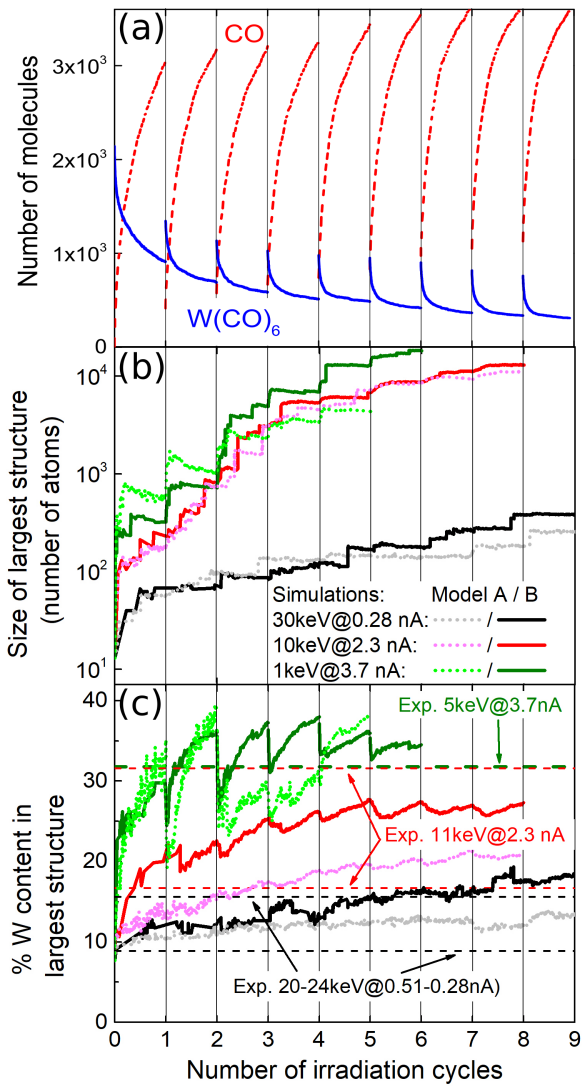}
	\caption{\label{fig3:Edep} 
		(a) Time evolution of the number of W(CO)$_6$ on SiO$_2$ (solid line) and free CO (dashed line) during FEBID with a 30keV@5.9 nA beam.
		(b) Evolution of the number of atoms and 
			(c) the W-metal content in the largest simulated islands for PE beams of energies 1, 10 and 30 keV, for different currents. Dotted and solid curves depict the results from two different chemistry models, in which dangling bonds within the same growing nanostructure are (model B) or are not (model A) allowed to recombine. Dashed horizontal lines, pointed by arrows, correspond to experimentally obtained compositions at the conditions indicated by the corresponding label \cite{Porrati2009}. 
	}
\end{figure}

Each irradiation phase lasts for a time known as dwell time, which typical values in experiment ($\geq \mu$s) are still computationally demanding for MD.
Instead, they are set here to 10 ns.
Consequently, simulated PE fluxes $J_0$ (and hence PE beam currents $I_0$) must be scaled to match the same number of PE per unit area and per dwell time as in experiments \cite{Sushko2016}
(see Supplementary Information S4.A).
As for replenishment, its characteristic times are also typically very long ($\sim$ms). In simulations, the CO molecules desorbed to the gas phase are simply removed during the replenishment stages and new W(CO)$_6$ molecules are deposited. 
Figure \ref{fig3:Edep}(a) illustrates these successive irradiation-replenishment stages by depicting the number of W(CO)$_6$ and free CO molecules during several 
of these cycles for a 30 keV PE beam of equivalent experimental current $I_0^{\rm exp} = 5.9$ nA (in short, 30keV@5.9nA). 

As the irradiation-replenishment cycles proceed, the process of nucleation of metal-enriched islands and its coalescence starts \cite{Sushko2016}. This is shown in Fig. \ref{fig3:Edep}(b), where the number of atoms in the largest 
island
is shown for three simulation conditions 
close to reported in experiments \cite{Porrati2009}: 30keV@0.28nA, 10keV@2.3nA and 1keV@3.7nA.
Results of two different models for the chemistry occurring within the growing nanostructure are presented \cite{Sushko2016}:
in model A (dotted lines), dangling bonds of a given nanostructure can only react with unsaturated bonds belonging to a different molecule; in model B (solid lines), the restructuring of bonds within a growing nanostructure is also allowed (see Supplementary Information S4.C for further details). The jumps in the size of the largest island observed occasionally are due to the merging of independent nanoclusters that grow on the substrate. 

As the islands grow, their average chemical composition also changes. The time evolution of the W-metal content of the largest island for the three 
aforementioned cases 
is depicted in Fig. \ref{fig3:Edep}(c) for the chemistry models 
A (dotted lines) and B (solid lines). The metal content grows fast during the first irradiation cycles, until it slowly starts to saturate for each set of beam parameters after $\sim4-$5 irradiation cycles. It is worth noting that our
simulation results are consistent with experimental data \cite{Porrati2009} for the 20--24keV@0.28--0.51nA, 11keV@2.3nA and 5keV@3.7nA cases, represented by dashed horizontal lines in Fig. \ref{fig3:Edep}(c).

\begin{figure}[t]
	\includegraphics[
	width=0.9\columnwidth
	]{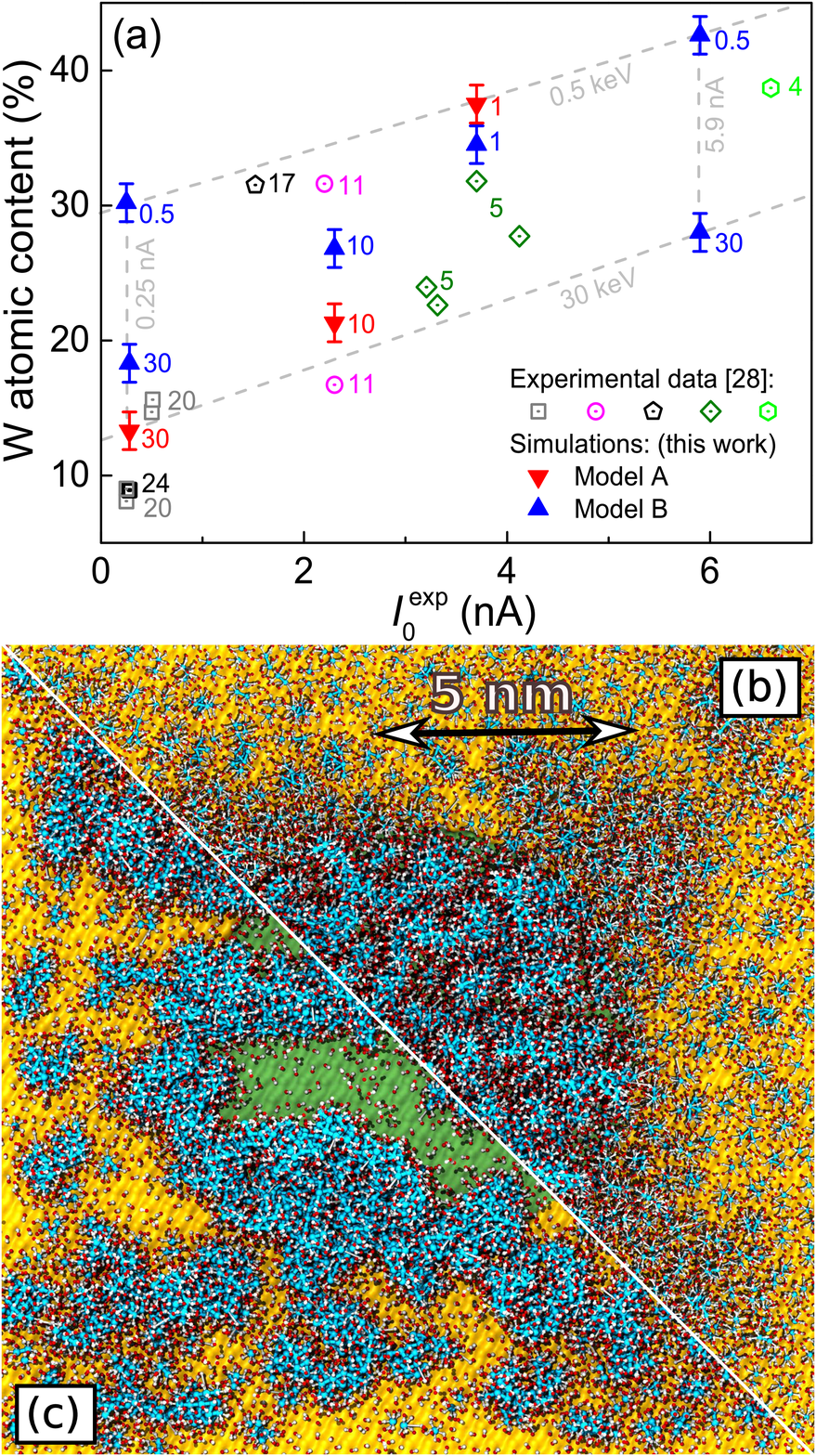}
	\caption{\label{fig6:TIW}
		(a) Dependence of the deposit metal content on the beam energy $T_0$ and current $I_0^{\rm exp}$, from experiments (open symbols) \cite{Porrati2009} and simulations (full symbols). Numbers next to symbols represent the beam energy in keV for each case. Lower panels show the top views of the deposits produced by (b) 10keV@2.3nA and (c) 1keV@3.7nA beams. The green area marks the PE beam spot while blue, white and red spheres represent, respectively, W, C and O atoms; the SiO$_2$ substrate is represented by a yellow surface.
	}
\end{figure}

Experimental measurements 
were limited to particular values of energy and current due to the characteristics of the electron source \cite{Porrati2009}. Nonetheless, our simulation method allows for the exploration of much wider regions of electron beam parameters. 
To do so, we also considered the cases of 30keV@5.9nA, 0.5keV@0.25nA and 0.5keV@5.9nA, obtaining the deposit metal contents depicted by full symbols in Fig. \ref{fig6:TIW}(a), as a function of experimentally equivalent current $I_0^{\rm exp}$. Error bars show the standard deviations obtained from three independent simulations for each case. 
Experimental results \cite{Porrati2009} are shown by open symbols.
Numbers next to symbols represent the beam energies in keV.
It is clearly seen that the results from simulations 
are within the range of experimental uncertainties,
which indicates the predictive capabilities of the simulations. 

The 
cases analysed in this investigation provide a detailed ``map'' of the attainable metal contents in the deposits as a function of the beam parameters, which is a very valuable outcome for the optimisation of FEBID 
with W(CO)$_6$@SiO$_2$. 
This is marked in Fig. \ref{fig6:TIW}(a) by dashed lines corresponding to the limiting values of energy and current studied.
These results 
clearly show that, within the analysed energy domain, a decrease in the beam energy and an increase in the current promote the faster growth of the deposit, as well as the augment in its metal content.
Simulation results provide the grounds for
clearly understanding such trends: an increment in the current means a larger number of PE per unit time, while a reduction in the energy produces an increase in the SE yield (Fig. \ref{fig1:yields}(a)). 
These lead to both the greater size of the deposit and its larger metal content due to the increased probability for bond cleavage (Figs. \ref{fig2:Nxy}(c)-(d)). It should be noted that a reduction of beam energy below $\sim$400 eV may diminish the metal content due to the lowering of the electron yields (Fig. \ref{fig1:yields}).

Finally, Figs. \ref{fig6:TIW}(b) and (c) show top views of the simulated deposits for 1keV@3.7nA and 10keV@2.3nA, after 5 and 7 irradiation cycles, respectively (the number of atoms in the largest island is similar in these cases, $\sim 12.000$). The central green circular surface marks the area covered by the PE beam (of 5 nm radius). 
These figures help to understand how different energy-current regimes can lead to distinct deposit microstructures. While the higher energy beam of 10 keV produces a deposit almost entirely localised within the intended nanomanufacturing region (i.e., the PE beam area), 
the lower energy beam of 1 keV produces a 
more sparse and ramified deposit (at least during the early stage of the FEBID process), that significantly extends beyond the PE beam area, 
producing an undesired edge broadening of the structure. 
Although a detailed analysis of these effects 
deserves a more in-depth analysis (which is not possible within the limits of the present manuscript), it is worth to note that the SE yield goes from  larger than 1 to lower than 1 in the 1--10 keV range (Fig. \ref{fig1:yields}(a)), SE being the main responsible for the 
beam halo (Fig. \ref{fig2:Nxy}). Such detailed predictions on the early stage of growth of metal deposits can be currently tested experimentally \cite{VanDorp2012a}. 

\section*{\label{sec:Conclusion}Conclusions}

In this study we have demonstrated how to couple detailed space and energy distributions of electrons at the substrate surface (obtained from MC calculations \cite{Dapor2017,DaporBook,Azzolini2018}) with radiation-induced dynamics and chemical reactions simulations (by means of the IDMD technique \cite{Sushko2016,Solovyov2017b}) 
in order to describe 
radiation effects at the 
molecular level for experimentally relevant timescales. As a particular case study, and due to its relevance in nanotechnology, we have analysed the FEBID process for W(CO)$_6$ precursor molecules on hydroxylated SiO$_2$. 

The presented results demonstrate how the novel MC-IDMD 
approach 
provides the necessary molecular insights into the key processes behind FEBID, which can be used for its further optimisation and development. Notably, the simulations (which rely on basic atomic and molecular data such as cross sections for electron scattering and molecular fragmentation) have demonstrated a great predictive power, yielding, for the first time, fabricated nanostructure compositions and morphologies in excellent agreement with available experimental data \cite{Porrati2009}.
Particularly, the increase in both the growth rate and W-metal content of the deposits with the increase in PE beam current and with the decrease in its energy, have been shown to be related to the increase in the number of ejected low energy SE. The latter are also responsible for the different microsrtuctures and edge broadenings observed for beams of different energies.
Many other aspects influencing FEBID and not addressed here (namely, other substrate-molecule combinations, different replenishment conditions \cite{Fowlkes2010}, the effects of contaminants or 
local heating by the PE beam \cite{Mutunga2019}, post-growth purification procedures...) can be analysed by utilising the protocols described in the present investigation.

Moreover, the new introduced methodology, which bridges the gap between other current approaches to describe radiation-induced effects spanning multiple space, time and energy scales, is general and 
readily 
applicable in many other important fields. 
It is worth noticing that mechanisms rather similar to the ones underlying FEBID (i.e., electron generation by different types of radiation, their transport and the chemistry induced on surfaces) are common to problems as diverse as the astrochemistry processes happening in interstellar ices due to cosmic radiation \cite{Tielens2013,Mason2014}, in the use of metallic nanoparticles as enhancers of modern radiotherapies 
\cite{Haume2016a,Solovyov2017} or in photoelectrochemical devices \cite{Zhang2016}.
This new MC-IDMD approach 
offers a valuable tool which might provide unprecedented insights in
many relevant problems in physics, chemistry, materials science, biomedicine and related technologies, in which irradiation-driven chemistry and multiscale phenomena play an essential role.

\section*{\label{sec:methods}Methods}


Simulations were performed by means of the irradiation driven MD (IDMD) method \cite{Sushko2016} implemented in the MBN (Meso-Bio-Nano) Explorer software package \cite{Solovyov2012,Solovyov2017b}.
Within this framework, the space-dependent rate for bond cleavage in molecules on the substrate surface is given by:
\begin{eqnarray}
P(x,y) & = & \sigma_{{\textrm{frag}}}(T_0) \, J_{\textrm{PE}}(x,y,T_0) + \nonumber \\ 
& \sum_i & \sigma_{{\textrm{frag}}}(T_i) \, J_{\textrm{SE/BSE}}(x,y,T_i) \, \mbox{,}
\label{eq:Pxy}
\end{eqnarray}
where a discrete set of values for the electron energies $T_i$ was assumed for simplicity, but without affecting the final results. 
$J_{\rm PE/SE/BSE}(x,y,T_i)$ are space- and energy-dependent 
fluxes of PE/SE/BSE (electrons per unit area and unit time) and $\sigma_{\rm frag}(T_i)$ is the energy-dependent molecular fragmentation cross section. 
The PE beam flux 
at the irradiated circular spot of radius $R$ is:
\begin{equation}
J_0 = \frac{I_0}{eS_0} \, \mbox{,}
\label{eq:J0}
\end{equation}
where $I_0$ corresponds to the PE beam current, $S_0 = \pi R^2$ to its area and $e$ is the elementary charge. 
Note that 
$\sum_i J_{\rm SE/BSE}(x,y,T_i) = J_{\rm SE/BSE}(x,y)$ 
gives the space-dependent fluxes which are plotted in Figs. \ref{fig2:Nxy}(a)-(b). For uniform PE beams, as used in this investigation, $J_{\rm PE}(x,y,T_0) = J_0$ for every point with coordinates $x^2+y^2 \leq R^2$.

The electron distributions were simulated using the MC radiation transport code SEED (Secondary Electron Energy Deposition) \cite{Dapor2017,DaporBook,Azzolini2018}. Molecular fragmentation and further chemical reactions were simulated by means of MBN Explorer \cite{Sushko2016,Solovyov2012,Solovyov2017b}. 
Its dedicated user interface and multi-task toolkit, MBN Studio \cite{Sushko2019}, was employed for constructing the molecular system, performing the precursor molecule replenishment phases, as well as for analysing the IDMD simulation results.

\subsection*{\label{sec:SEED}Monte Carlo code SEED}

The SEED 
code follows the classical trajectories of energetic electrons travelling inside a 
condensed phase material, by employing the usual Monte Carlo recipes for electron transport simulation \cite{DaporBook,Dapor2017,Azzolini2018}. It is based on the calculation of (i) the differential inelastic scattering cross sections accurately obtained by using the dielectric formalism \cite{Dapor2017,deVera2013,deVera2015}, (ii) the electron-phonon quasi-elastic scattering cross-section computed by the use of the Fr\"ohlich theory \cite{Frohlich1954} and (iii) the differential elastic scattering cross section performed by the relativistic partial wave expansion method (RPWEM) \cite{Jablonski2004} including the Ganachaud and Mokrani empirical correction for low electron energies ($\leq$ 20--30 eV) \cite{Ganachaud1995}.

The empirical parameters for the Fr\"{o}hlich and the Ganachaud-Mokrani theories are set in order to reproduce by simulation the experimentally known SE yield for SiO$_2$ \cite{Glavatskikh2001,Yi2001}.
See Supplementary Information S2 and Refs. \cite{DaporBook,Dapor2017,Azzolini2018} for extended discussions on the SEED code and its validation.

\subsection*{\label{sec:MBN}MBN Explorer and Irradiation Driven Molecular Dynamics}

MBN 
Explorer is a multi-purpose software package for advanced multiscale simulations of structure and dynamics of complex molecular systems \cite{Solovyov2012,Solovyov2017b}, featuring a wide variety of computational algorithms for the simulation of atomistic and coarse-grained systems. 
It includes the advanced algorithms of reactive MD \cite{Sushko2015} and the unique IDMD \cite{Sushko2016} exploited in this investigation.

In the MD approach, the dynamics of a system is followed
by numerically solving the coupled classical Langevin equations of motion of all its constituent atoms. 
The interaction forces are treated in this work by means of the CHARMM force field \cite{MacKerell1998}. 

The IDMD algorithm implemented in MBN Explorer \cite{Sushko2016} superimposes random processes of molecular bond breakage due to irradiation during classical reactive MD. These processes are treated as local (involving the atoms participating in a chemical bond) energy deposition events occurring in the sub-femtosecond timescale, so they are considered to happen instantaneously between successive simulation time steps. They occur randomly, with a rate determined by the probabilities for quantum processes such as dissociative ionisation or dissociative electron attachment, Eq. (\ref{eq:Pxy}). The fast relaxation of the excess energy after these interactions results in the cleavage of particular bonds and the formation of active species (radicals with unsaturated dangling bonds) which can undergo further chemical reactions. 

The cleavage  
and formation of chemical bonds 
and the monitoring of the system's dynamical topology, along with the redistribution of atomic partial charges, is managed by means of the reactive version of the CHARMM force field implemented in MBN Explorer 
\cite{Sushko2015}. Its parameterisation for the W(CO)$_6$ molecule was described in an earlier study \cite{deVera2019frag}.
In this investigation we assume that every fragmentation event leads to the cleavage of a single W-C bond, while the much stronger C-O bonds will not react \cite{deVera2019frag}. The energy deposited in the cleaved W-C bonds is chosen in accordance with average values obtained from mass spectrometry experiments \cite{Cooks1990,Wnorowski2012} and dedicated simulations of the molecule fragmentation \cite{deVera2019frag}, see Supplementary Information S4.B.

The details of the IDMD methodology are explained in \cite{Sushko2016}, and all necessary details for its application to the system studied in this investigation are given in Supplementary Information S4.

\begin{acknowledgments}
	PdV gratefully acknowledges the Alexander von Humboldt Foundation/Stiftung and 
	the Spanish Ministerio de Ciencia e Innovaci\'{o}n for their financial support by means of, respectively, Humboldt (1197139) and Juan de la Cierva (FJCI-2017-32233) postdoctoral fellowships. MA is thankful to Prof. Nicola M. Pugno for managing her financial support. IAS acknowledges the Lundbeck Foundation and the Volkswagen Foundation (Lichtenberg Professorship) for their support. This work was also supported in part by the Deutsche Forschungsgemeinschaft
	(Projects no. 415716638 and GRK1885), the Spanish Ministerio de Ciencia e
	Innovaci\'{o}n and the European Regional Development
	Fund (Project no. PGC2018-096788-B-I00), by the Fundaci\'{o}n S\'{e}neca -- Agencia de Ciencia y Tecnolog\'{i}a de la Regi\'{o}n de Murcia (Project No. 19907/GERM/15), by the Conselleria d'Educaci\'{o}, Investigaci\'{o}, Cultura i Esport de la Generalitat Valenciana (Project no. AICO/2019/070) and by the
	COST Action CA17126 ``Towards understanding and modelling
	intense electronic excitation'' (TUMIEE). The possibility
	to perform computer simulations at Goethe-HLR 
	cluster of the Frankfurt Center for Scientific Computing is
	gratefully acknowledged.
\end{acknowledgments}


\section*{Additional information}
Supplementary information is available in the online version of the manuscript. 

 

\bibliography{library}
\bibliographystyle{naturemag}

\end{document}


\title{Multiscale simulation of the focused electron beam induced deposition process \\
(Supplementary Information)}

\author{Pablo de Vera$^{1,2}$}
\email{pablo.vera@um.es}
\author{Martina Azzolini$^3$}
\author{Gennady Sushko$^1$}%
\author{Isabel Abril$^4$}
\author{Rafael Garcia-Molina$^2$}
\author{Maurizio Dapor$^3$}
\author{Ilia A. Solov'yov$^{5}$}
\email{On leave from A. F. Ioffe Physical-Technical Institute, 194021 St. Petersburg, Russia}
\author{Andrey V. Solov'yov$^{1}$}%
\email{On leave from A. F. Ioffe Physical-Technical Institute, 194021 St. Petersburg, Russia}
\affiliation{%
	$^1$MBN Research Center, Altenh\"{o}ferallee 3, 60438 Frankfurt am Main, Germany
}%
\affiliation{$^2$Departamento de F\'{i}sica -- Centro de Investigaci\'{o}n en \'{O}ptica y Nanof\'{i}sica (CIOyN), Universidad de Murcia, E-30100 Murcia, Spain
}%
\affiliation{$^3$European Centre for Theoretical Studies in Nuclear Physics and Related Areas (ECT*), I-38123 Trento, Italy
}%
\affiliation{$^4$Departament de F\'{i}sica Aplicada, Universitat d'Alacant, E-03080 Alacant, Spain
}%
\affiliation{%
	$^5$Department of Physics, Carl von Ossietzky University, Carl-von-Ossietzky Stra\ss e 9-11, 26129 Oldenburg, Germany
}%

\date{\today}


\maketitle

\newpage

\section{Molecular adsorption energy and simulation of surface diffusion}

The diffusion coefficient of molecules on a surface strongly depends on temperature ${\cal T}$ and the activation energy for diffusion $E_{\rm diff}$
\cite{Utke2008}.
The latter is related to the adsorption energy $E_{\rm ads}$ for the molecule on the substrate, leading to the following relation:
\begin{equation}
D({\cal T},E_{\rm ads}) = D_0 \exp{\left(-\frac{E_{\rm ads}}{k{\cal T}}\right)} \mbox{,}
\label{eq_DvsT2}
\end{equation}
where $D_0$ is the diffusion coefficient at very large temperatures and $k$ is the Boltzmann's constant.
The adsorption energy $E_{\rm ads}$ depends on the strength of the 
adhesion
forces between the molecule and the substrate. 

The molecule-substrate forces (arising in our simulations from van der Waals interactions 
between pairs of atoms) are modelled in molecular dynamics (MD)
by means of the
Lennard-Jones potential, which needs two parameters (interatomic distance
and depth
of the potential well). 
For W, C, Si, O and H atoms, different sets of van der Waals parameters can be found in the literature \cite{Sushko2016,Athanasopoulos1992,Bernardes2013,Filippova2015}.
In the present study,
these parameters have been combined in different ways, producing several sets which allow screening a wide region of potential adsorption energies, 
until 
reaching a  diffusion coefficient in reasonable agreement with the 
only available (to the best of our knowledge) empirical information 
(estimated from the adjustment of model calculations to experimental data on nanostructure growth) 
\cite{Fowlkes2010}. 
Note that the choice could have been also performed relying on \textit{ab initio} calculations of the adsorption energy, although we preferred to rely here on empirical information.

\begin{figure}[t]
	\centering
	\includegraphics[width=0.6\columnwidth]{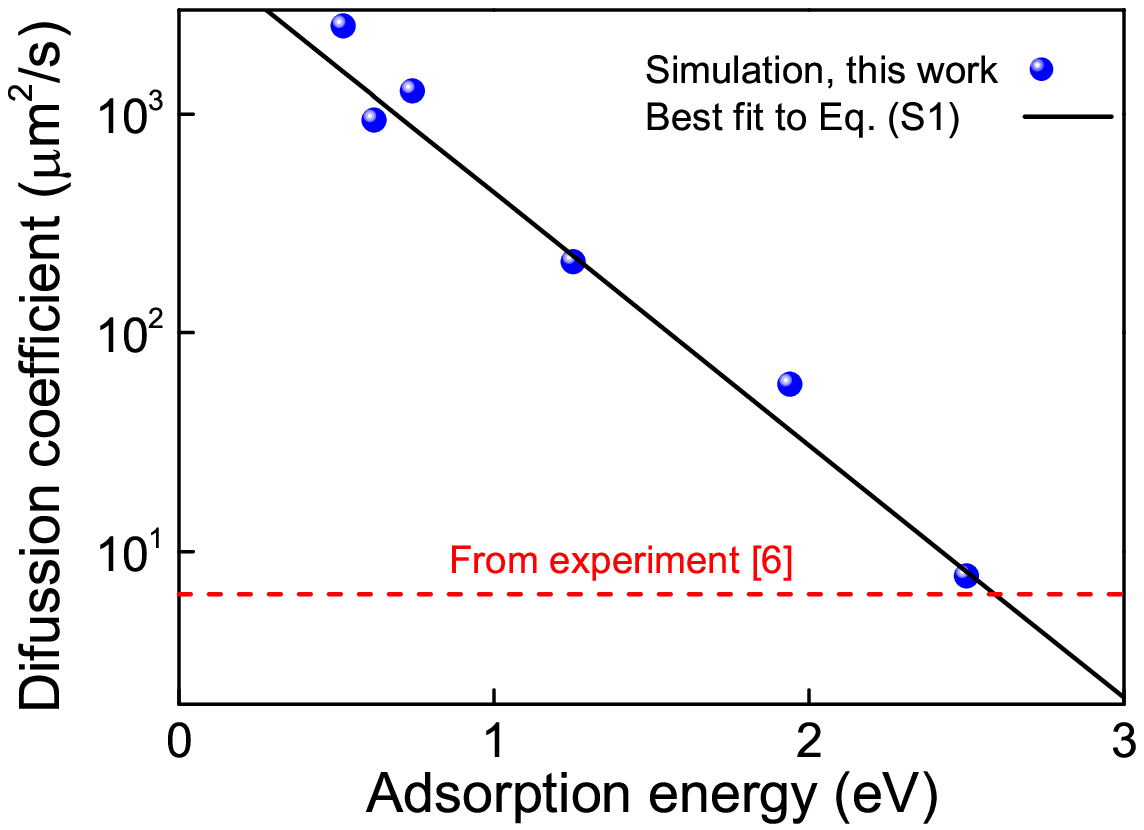}
	\caption{\label{fig1:Diff}
		W(CO)$_6$ surface diffusion coefficient on hydroxylated SiO$_2$ as a function of the adsorption energy for different sets of van der Waals parameters. An empirically estimated value 
	(the adsorption energy was not determined) is shown by a dashed line \cite{Fowlkes2010}.
	}
\end{figure}

The diffusion coefficients and adsorption energies of
W(CO)$_6$ molecules on hydroxylated SiO$_2$ 
have been obtained 
from MD simulations of 2 ns duration at ${\cal T} = $ 300 K for each set of parameters, as explained in \cite{Sushko2016}. The results are depicted in Fig. \ref{fig1:Diff} by symbols; the solid line represents the best fit by means of Eq. (\ref{eq_DvsT2}). The experimentally determined value 
\cite{Fowlkes2010} is shown by a dashed line (the adsorption energy was not empirically obtained). An adsorption energy 
around 2.5 eV 
yields a
surface diffusion coefficient
of 7.71 $\mu$m$^2$/s, fairly close to the experimentally determined value of 6.64 $\mu$m$^2$/s \cite{Fowlkes2010}. 
Parameters for C, H and O atoms come from  \cite{Sushko2016}, those for Si from  \cite{Athanasopoulos1992}, while for W they come from  \cite{Filippova2015}.

\section{Monte Carlo code SEED for electron transport and interaction cross sections for silica}

The details of the SEED (Secondary Electron Energy Deposition) Monte Carlo (MC) code for energetic electron transport in solids are explained in detail in  \cite{DaporBook}.
It is based on the calculation of (i) the differential inelastic scattering cross sections obtained by using the dielectric formalism \cite{
deVera2013,deVera2019}, (ii) the electron-phonon quasi-elastic scattering cross-sections computed by the use of the Fr\"ohlich theory \cite{Frohlich1954} and (iii) the differential elastic scattering cross sections performed by the relativistic partial wave expansion method (RPWEM) \cite{Jablonski2004} including the Ganachaud and Mokrani empirical correction for low electron energies ($\leq$ 20--30 eV) \cite{Ganachaud1995}. 
It should be noted that coherent scattering effects due to the long De Broglie wavelength for very low energy electrons are not yet considered in any MC code. Nonetheless, the good comparison of the results obtained by means of SEED with experimental and reference data (see below or \cite{DaporBook} for further examples) demonstrates that its accuracy is good enough for the purposes of the present investigation.

\begin{figure}[t]
	\centering
	\includegraphics[width=1.0\columnwidth]{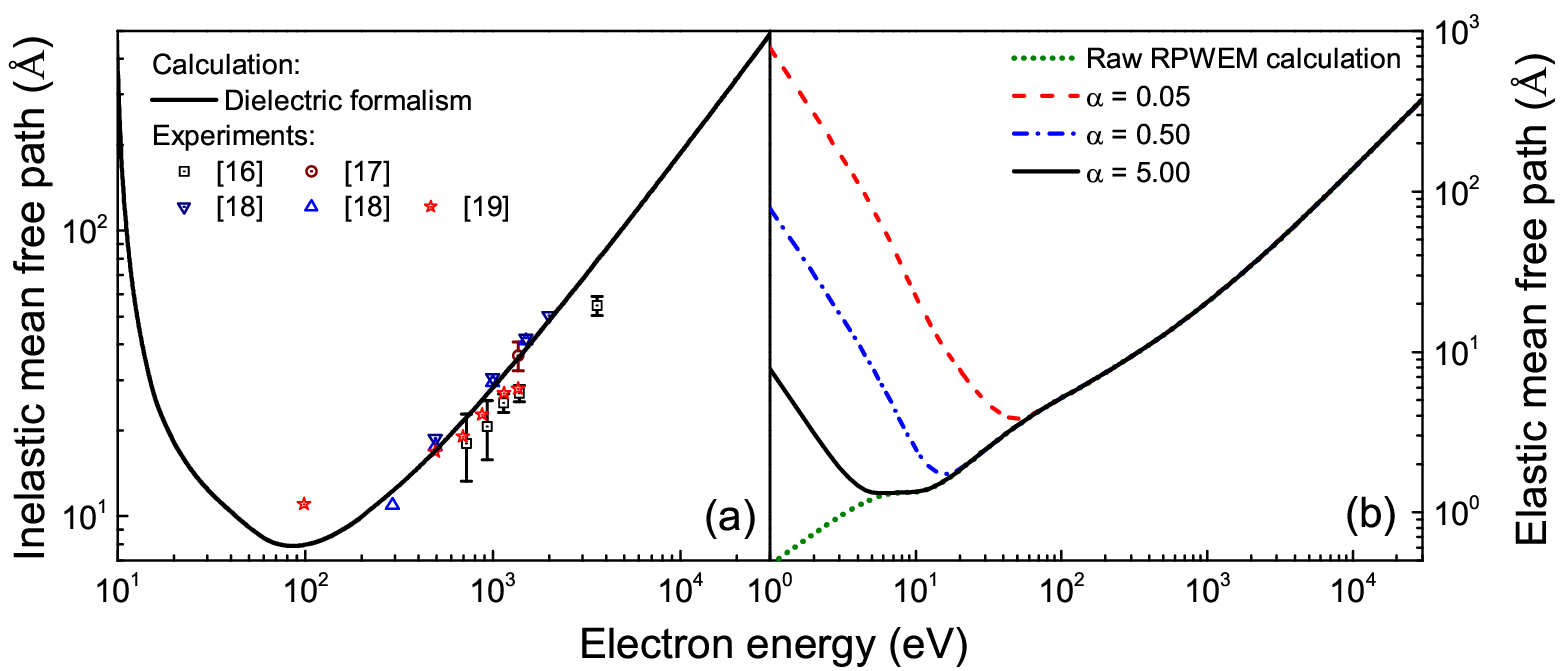}
	\caption{\label{fig2:SiO2XS}
	(a) Calculated inelastic mean free path for electrons in SiO$_2$ (solid line) compared to  
	experimental data 
	(symbols) \cite{Flitsch1975,Hill1976,Jung2003,Murat2004}.
	(b) Calculated elastic mean free path for electrons in SiO$_2$: the dotted line depicts the raw RPWEM calculation, while other lines show the results obtained with 
	different choices of the Ganachaud-Mokrani parameter $\alpha$.}
\end{figure}

The dielectric formalism, very successful for the calculation of electronic interaction cross sections in condensed matter, requires the knowledge of the energy-loss function (ELF) of the target material, Im$\left[-1/\epsilon(k,\omega)\right]$ (where $\epsilon(k,\omega)$ is the complex dielectric function), accounting for its electronic excitation spectrum for excitations of given energy and momentum, $\hbar\omega$ and $\hbar k$, respectively. 
An estimate for the mean binding energy of the outer-shell electrons of the target is needed in order to disentangle the processes of ionisation and electronic excitation \cite{deVera2013}. The optical ELF ($\hbar k =0$) of SiO$_2$ (mass density 2.19 g/cm$^3$) has been taken from the compilation of  \cite{Palik1999}
and extended to finite momentum transfers by means of the Mermin Energy Loss Function-Generalised Oscillator Strengths (MELF-GOS) method \cite{Garcia-Molina2011},
tested for many condensed-phase materials. A mean binding energy of 12.2 eV has been estimated from data taken from  \cite{
	Garvie1998}. 
The effect of surface hydroxylation in the electronic properties of SiO$_2$ is not deemed to affect much the electron propagation through the substrate, which mostly occurs within the bulk of the material.
The calculated total inelastic mean free path 
is shown by a solid line in Fig. \ref{fig2:SiO2XS}(a) 
and compared 
to the available experimental data 
\cite{Flitsch1975,Hill1976,Jung2003,Murat2004}, finding a
rather good agreement
in a wide energy range.

\begin{figure}[b]
	\centering
	\includegraphics[width=1.0\columnwidth]{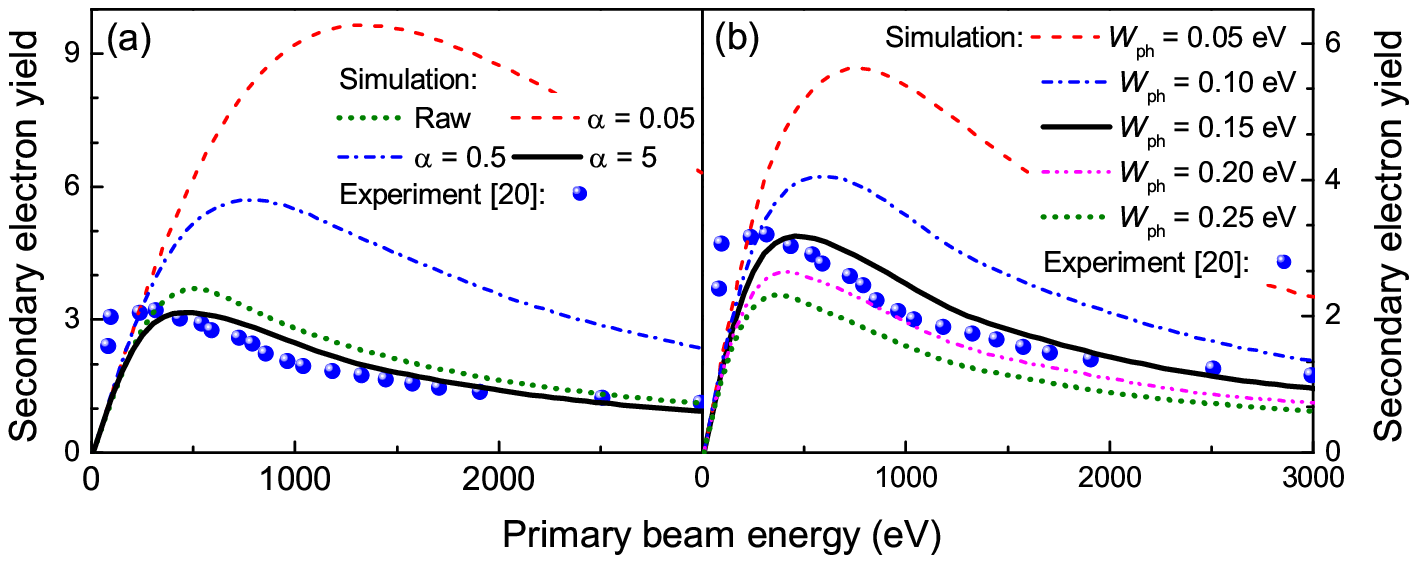}
	\caption{\label{fig3:MCparameters} 
	Calculated secondary electron yield (a) for different values of the Ganachaud-Mokrani parameter $\alpha$ (lines), and (b) for different values of the phonon energy $W_{\rm ph}$, as compared to the experimental values (symbols) \cite{Glavatskikh2001}.
    }
\end{figure}

The elastic mean free path for electrons in SiO$_2$ calculated by meas of the RPWEM is shown in Fig. \ref{fig2:SiO2XS}(b) (dotted line). This method is known to provide unreasonable small values 
(shorter than interatomic distances)
for $T<10$ eV. 
To amend this tendency, the Ganachaud and Mokrani empirical correction is used \cite{Ganachaud1995}, which requires the setting of the Ganachaud-Mokrani 
parameter $\alpha$. Figure \ref{fig2:SiO2XS}(b) shows the effect of different choices of $\alpha$ in the resulting elastic mean free path. 
Furthermore, the Fr\"ohlich theory \cite{Frohlich1954} for electron-phonon scattering requires the knowledge of the characteristic phonon energy, $W_{\rm ph}$, whose typical values are in the range $ 0.01-0.1$ eV. 

All elastic, inelastic and electron-phonon interactions have an impact in the trajectories of (especially low energy) electrons, affecting their ability to escape from the substrate and hence determining the secondary electron (SE) yield. Since inelastic and  elastic (high energy) cross sections are fully determined
(the former from the dielectric theory and the latter from the RPWEM calculations), 
only the setting of the free parameters for the Ganachaud-Mokrani and Fr\"{o}hlich theories remains.
These parameters have been set in order to correctly simulate the experimentally known SE yield for SiO$_2$ \cite{Glavatskikh2001}.
Figure \ref{fig3:MCparameters}(a) shows the effect of different choices of the Ganachaud-Mokrani parameter 
in the simulated SE yield (assuming $W_{\rm ph} = 0.15$ eV),
while Fig. \ref{fig3:MCparameters}(b) shows
the effect of different phonon energies (fixing $\alpha = 5$).
Values of $\alpha = 5$ and $W_{\rm ph} = 0.15$ eV (solid lines) provide a very reasonable agreement with the experimental data (symbols) \cite{Glavatskikh2001}.

\begin{figure}[t]
	\includegraphics[width=0.7\columnwidth]{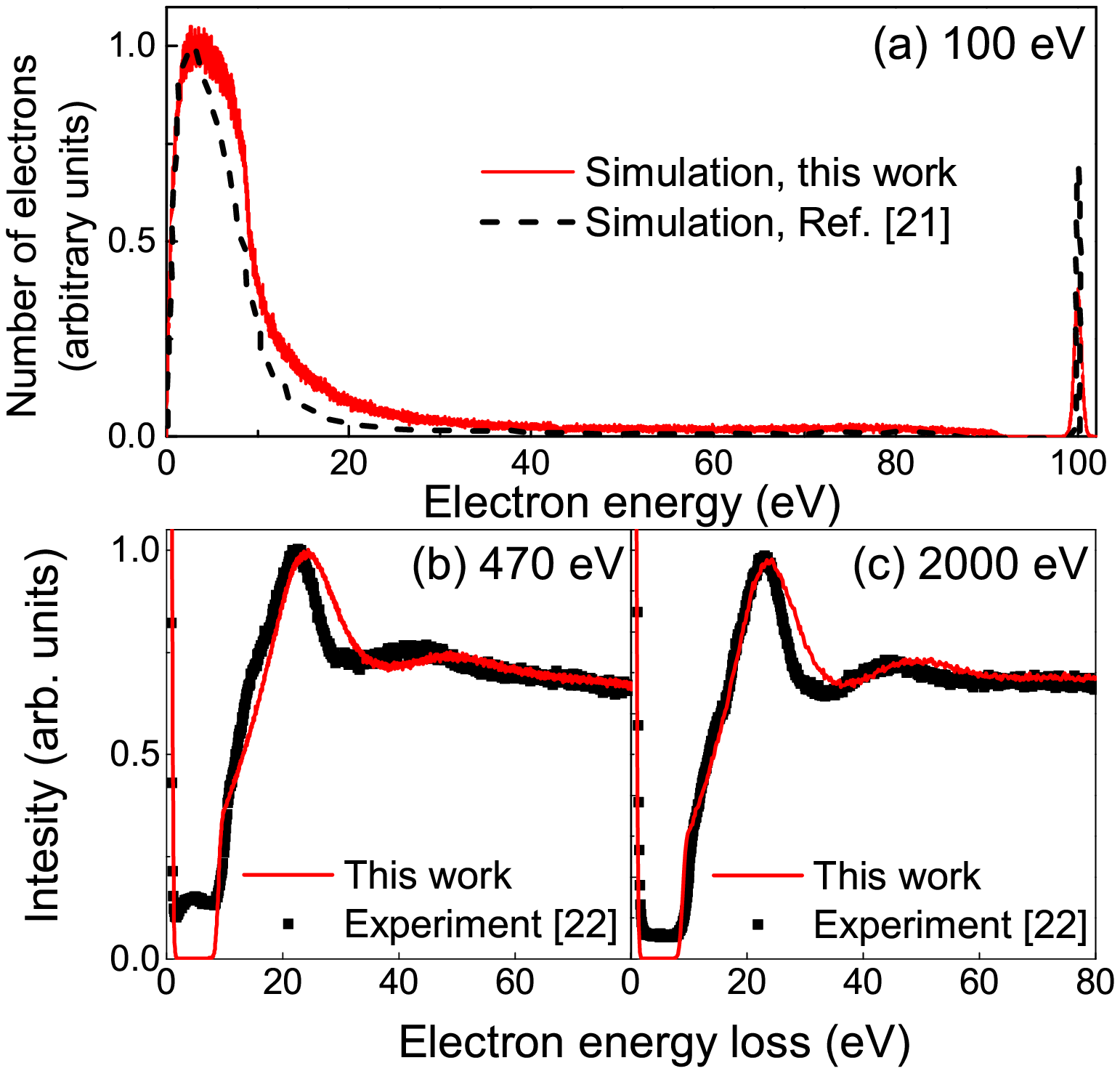}
	\caption{\label{fig3:REELS} 
		(a) Simulated SE and BSE energy spectrum for a 100 eV PE beam impinging in SiO$_2$: present calculation (solid histogram) and simulations from  \cite{Schreiber2002} (dashed histogram). The other panels depict simulated reflection electron energy loss spectra (REELS, lines) for PE beams of (b) 470 eV and (c) 2 keV, compared to experiments (symbols) \cite{Filippi2007}.}
\end{figure}

The reliability of the MC simulations has been tested by calculating the energy spectra as well as energy loss spectra of the reflected electrons (i.e., the reflection electron energy loss spectra, REELS) and 
comparing them 
to reference and experimental data. Figure \ref{fig3:REELS}(a) compares the calculated (solid histogram) energy spectrum of ejected SE and backscattered electrons (BSE) for a 100 eV primary electron (PE) beam impinging in SiO$_2$ with reference simulations (dashed histogram) 
\cite{Schreiber2002}. Figures \ref{fig3:REELS}(b) and (c) compare the simulated REELS spectra (red lines) with experimental data (black symbols) \cite{Filippi2007} for PE beams of 470 eV and 2000 eV, respectively. Simulations qualitatively agree with the energy spectrum from \cite{Schreiber2002} and are in an excellent agreement with the experimental REELS spectra \cite{Filippi2007}.

\section{W(CO)$_6$ molecule fragmentation cross sections}

For the W(CO)$_6$ molecule, 
experimental measurements exist for the \textit{relative} cross sections for electron-impact dissociative ionisation (DI) \cite{Wnorowski2012a} and for lower energy channels \cite{Wnorowski2012}, 
although no information about the \textit{absolute} cross sections
is available. 

The experimental data on DI relative cross sections reveals that almost every ionising collision ($\ge$ 97\%) leads to W(CO)$_6$ fragmentation to some extent. Thus, it is possible to approximate the total DI cross section, $\sigma_{\rm DI}(T)$, by estimating the total ionisation cross section employing the dielectric formalism \cite{deVera2013,deVera2019}  (see Supplementary Information S2). The optical ELF of W(CO)$_6$ has been estimated from a parametric approach for organic materials \cite{deVera2013,Tan2004} and a binding energy for the outer-shell electrons of 8.47 eV has been estimated from  \cite{Wnorowski2012a}. The calculated DI cross section is shown in Fig. 1(c) of the main text.

The absolute value of the lower energy dissociation channels cross section, $\sigma_{{\rm low-}T}(T)$,
can be fixed by calculating the experimentally known decomposition cross section 
for W(CO)$_6$ molecules on SiO$_2$ by a PE beam of energy $T_0$:
\begin{equation}
\sigma_{\rm decomp}(T_0) = \frac{1}{N_{\rm PE}}\int_0^{T_0} \frac{{\rm d}N(T)}{{\rm d}T}\sigma_{\rm frag}(T){\rm d}T \, \mbox{,}
\label{eq:averXS}
\end{equation}
where ${\rm d}N(T)/{\rm d}T$ is the energy spectrum of PE and the generated SE and BSE crossing the surface near the beam area (further normalised to the number $N_{\rm PE}$ of PE) and $\sigma_{\rm frag}(T) = \sigma_{\rm DI}(T)+\sigma_{{\rm low-}T}(T)$. The decomposition cross section has been experimentally determined to be 1.2 \AA$^2$ for 30 keV PE \cite{Hoyle1994}.
The energy spectrum is obtained from MC simulations (see Fig. 1(b) of the main text).
The experimental relative 
low energy dissociation cross section $\sigma_{{\rm low-}T}(T)$ \cite{Wnorowski2012} has been scaled so the integral of Eq. (\ref{eq:averXS}) gives a value of 1.2 \AA$^2$.

\begin{figure}[t]
\includegraphics[width=0.6\columnwidth]{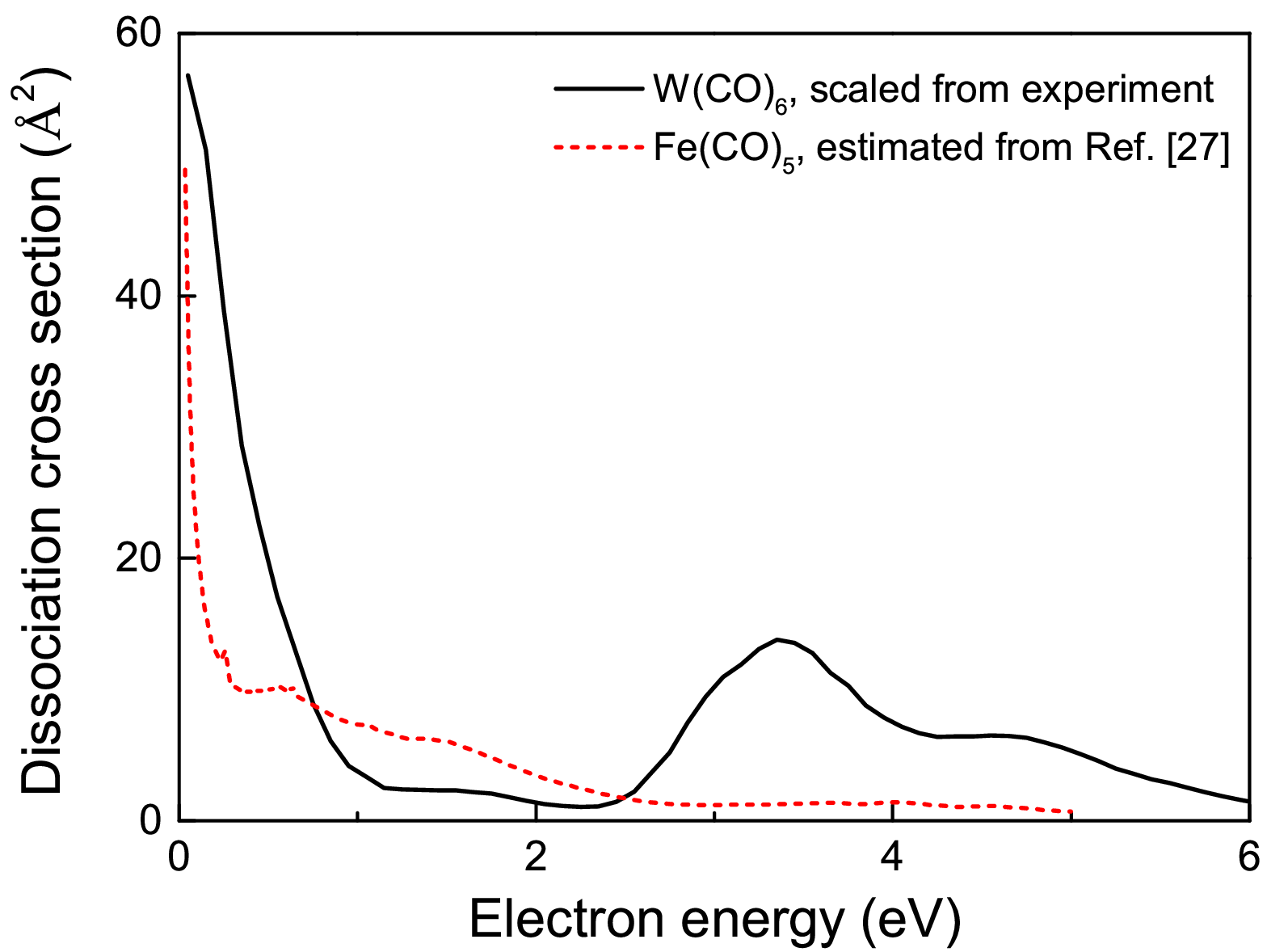}
\caption{\label{fig2:DEAXS}
	Comparison of the scaled low energy dissociation cross section for W(CO)$_6$ (this work) with that of Fe(CO)$_5$ (estimated from experiments and calculations from  \cite{Allan2018}).}
\end{figure}

The resulting 
low energy 
cross section 
is shown by a solid line in Fig. \ref{fig2:DEAXS}, and is compared with 
that
for the 
similar metal carbonyl Fe(CO)$_5$. For the latter molecule, measurements of the relative 
cross sections also exist, as well as calculations of its absolute values 
\cite{Allan2018}, which allow estimating the absolute 
cross section (dashed line in Fig. \ref{fig2:DEAXS}). 
The scaled 
cross section for W(CO)$_6$ 
is comparable with
the absolute cross section for Fe(CO)$_5$,
presenting the main characteristic peak (corresponding to the loss of a single CO ligand \cite{Wnorowski2012}), of a similar height, at $T \sim 0$ eV. 
Other dissociation channels observed for W(CO)$_6$ at 3.3 and 4.7 eV (corresponding to the loss of two and three CO ligands, respectively \cite{Wnorowski2012}) seem to be much weaker in Fe(CO)$_5$. 
The good comparison validates the scaling procedure employed.

\section{Irradiation driven molecular dynamics simulations of the FEBID process with MBN Explorer}

The fundamentals of irradiation driven molecular dynamics (IDMD) are summarised in ``Methods'' and full details can be found in \cite{Sushko2016}. Here, the main aspects influencing simulations within the current investigation are explained.

\subsection{Scaling of primary electron beam currents in simulations}

The molecular fragmentation rate is influenced (see ``Methods'') by the PE flux, i.e., the number $N_{\rm PE}$ of delivered PE per unit time $t$ and unit area $S$ at the irradiated spot:
\begin{equation}
J_0 = \frac{I_0}{eS_0} \, \mbox{,}
\label{eq_J0}
\end{equation}
where $I_0 = e\,{\rm d}N_{\rm PE}/{\rm d}t$ is the PE beam current, 
$S_0=\pi R^2$ is the PE beam area (with $R$ being the beam radius) and $e$ the elementary charge. Typical FEBID experiments use a beam radius of several nanometres. Here, 
it was set to $R=5$ nm. For simplicity, uniform PE fluxes $J_0$ were simulated by the MC code SEED (Supplementary Information S2) within the beam area $S_0$. In all MBN Explorer simulations, the 20$\times$20 nm$^2$ substrate is covered by 1--2 layers of W(CO)$_6$ molecules (density 5.4 molecules/nm$^2$),
which guarantees the full coverage of the substrate while keeping a layer thin enough to not significantly affect PE trajectories.

Currents of the order $I_0 \sim$ pA--nA for irradiation (dwell) times $\tau_{\rm d} \sim$ $\mu$s or longer are commonly used in experiments. However, currently it would be computationally very costly to perform MD simulations for such long times. Instead, in this work a number of PE per unit area and per dwell time similar to experiments was sought. For this purpose, the simulated currents (and hence the fluxes) are scaled in the following manner \cite{Sushko2016}:
\begin{equation}
\frac{I_0^{\rm sim}}{I_0^{\rm exp}} = \lambda \frac{S_0^{\rm sim}}{S_0^{\rm exp}} = \lambda \frac{R^{{\rm sim}\,2}}{R^{{\rm exp}\,2}} \, \mbox{,}
\end{equation}
where $\lambda = \tau_{\rm d}^{\rm exp}/\tau_{\rm d}^{\rm sim}$;
the super-indexes ``sim'' refer to simulation beam parameters while ``exp'' to experimental beam parameters. In such a way, simulated beam currents $I_0^{\rm sim} \sim \mu$A give the same amount of PE per simulated dwell time $\tau_{\rm d}^{\rm sim} \sim$ ns (which is feasible for MD simulations) as experimental beam currents $I_0^{\rm exp} \sim$ nA for experimental dwell times $\tau_{\rm d}^{\rm exp} \sim \mu$s. In Table \ref{table:Isim}, the PE beam parameters corresponding to each simulation (this work) and each experiment from  \cite{Porrati2009} (to which we compare) are summarised.

\begin{table}[t]
	\centering
	\caption{Beam parameters corresponding to each simulation (present work) and to each experiment reported in  \cite{Porrati2009}. Simulations and experiments marked in bold were performed with similar PE fluxes.}
	\begin{tabular}{|$c||^c|^c|^c|^c|^c|^c|}
		\hline 
		Simulation & $T_0$ & $R$ & $\tau_{\rm d}$ & $I_0^{\rm sim}$ & $I_0^{\rm exp}$ & PE flux 
		\\ 
		 & (keV) & (nm) & (ns) & ($\mu$A) & (nA) & (elec./nm$^2$/$\tau_{\rm d}$) \\ 
		\hline \hline
		0.5keV@0.25nA & 0.5 & 5 & 10 & 0.625 & 0.25 & 500 \\ 
		\hline
		0.5keV@5.9nA & 0.5 & 5 & 10 & 14.75 & 5.9 & 11700 \\ 
		\hline
		\rowstyle{\bfseries} 1keV@3.7nA & 1 & 5 & 10 & 9.25 & 3.7 & 7352 \\ 
		\hline
		\rowstyle{\bfseries} 10keV@2.3nA & 10 & 5 & 10 & 5.75 & 2.3 & 4570 \\ 
		\hline 
		\rowstyle{\bfseries} 30keV@0.28nA & 30 & 5 & 10 & 0.7 & 0.28 & 556 \\ 
		\hline
		30keV@5.9nA & 30 & 5 & 10 & 14.75 & 5.9 & 11700 \\ 
		\hline \hline
		Experiment & $T_0$ & $R$ & $\tau_{\rm d}$ & $I_0^{\rm sim}$ & $I_0^{\rm exp}$ & PE flux 
		\\ 
		& (keV) & (nm) & ($\mu$s) & ($\mu$A) & (nA) & (elec./nm$^2$/$\tau_{\rm d}$) \\ 
		\hline \hline
		4keV@6.6nA & 4 & 10 & 5 & - & 6.6 & 656 \\ 
		\hline
		5keV@3.2nA & 5 & 10 & 100 & - & 3.2 & 6371 \\ 
		\hline
		5keV@3.3nA & 5 & 10 & 100 & - & 3.3 & 6589 \\ 
		\hline
		\rowstyle{\bfseries} 5keV@3.7nA & 5 & 10 & 100 & - & 3.7  & 7352 \\ 
		\hline
		5keV@4.1nA & 5 & 10 & 100 & - & 4.1 & 8189 \\ 
		\hline
		\rowstyle{\bfseries} 11keV@2.3nA & 11 & 10 & 100 & - & 2.3 & 4570  \\ 
		\hline
		17keV@1.52nA & 17 & 10 & 100 & - & 1.52 & 3020  \\ 
		\hline
		\rowstyle{\bfseries} 20keV@0.25nA & 20 & 10 & 100 & - & 0.25 & 497 \\ 
		\hline
		20keV@0.5nA & 20 & 10 & 100 & - & 0.5 & 994 \\ 
		\hline
		20keV@0.51nA & 20 & 10 & 100 & - & 0.51 & 1013 \\ 
		\hline
		\rowstyle{\bfseries} 24keV@0.28nA & 24 & 10 & 100 & - & 0.28 & 556 \\ 
		\hline
	\end{tabular} 
	\label{table:Isim}
\end{table}

\subsection{Energy deposited to fragmenting bonds}

Bond dissociation events
are simulated by the deposition of an average energy per electron-molecule collision, $E_{\rm dep}$, in the atoms forming the bond, so that the atomic velocities increase fulfilling the requirements of momentum conservation \cite{deVera2019frag}. For the sake of simplicity, in this work it is assumed that every fragmentation event leads to the cleavage of a single W-C bond, while the much stronger C-O bonds will stay intact \cite{deVera2019frag}. Further collisions of the 
fragments with the environment may lead to subsequent cleavage of additional W-C bonds \cite{deVera2019frag}. 

The choice of $E_{\rm dep}$ influences the kinetics of the electron-driven chemical reactions, since low values favour metal-ligand recombination after dissociation, while larger values effectively put the metal atom and the ligand apart.
A value of $E_{\rm dep} = 325$ kcal/mol ($\sim$ 14 eV) has been chosen in simulations, since it is consistent with average values obtained from mass spectrometry experiments \cite{Cooks1990,Wnorowski2012} and dedicated W(CO)$_6$ fragmentation simulations \cite{deVera2019frag}.

Unfortunately, the kinetics of W(CO)$_6$ molecules fragmentation has not been measured on SiO$_2$ substrates, although it is known for gold substrates \cite{Rosenberg2013}. In \cite{Rosenberg2013}, 1--2 layers of W(CO)$_6$ molecules were deposited on Au, as in present simulations in SiO$_2$, and the effective decomposition cross section $\sigma_{\rm decomp}(T_0)$ (see Supplementary Information S3) was measured for a $T_0 = 500$ eV broad PE beam. On the one hand, this dense packing of molecules on the substrate prevents surface diffusion to some extent. On the other hand, the broad macroscopic beam (of $\sim$ 1 cm$^2$ area \cite{Rosenberg2013}) characterised by a decomposition cross section $\sigma_{\rm decomp}(T_0)$ produces a uniform fragmentation probability $P = J_0 \, \sigma_{\rm decomp}(T_0)$ over a nanometric area. As a consequence, it might be possible to, at least qualitatively, compare simulations on SiO$_2$ to the reported experiments on a Au substrate \cite{Rosenberg2013}, despite their potential differences in terms of surface diffusion and electron emission, provided that the empirical decomposition cross section is employed for these specific simulations.

\begin{figure}[t]
	\centering
	\includegraphics[width=0.6\columnwidth
	]{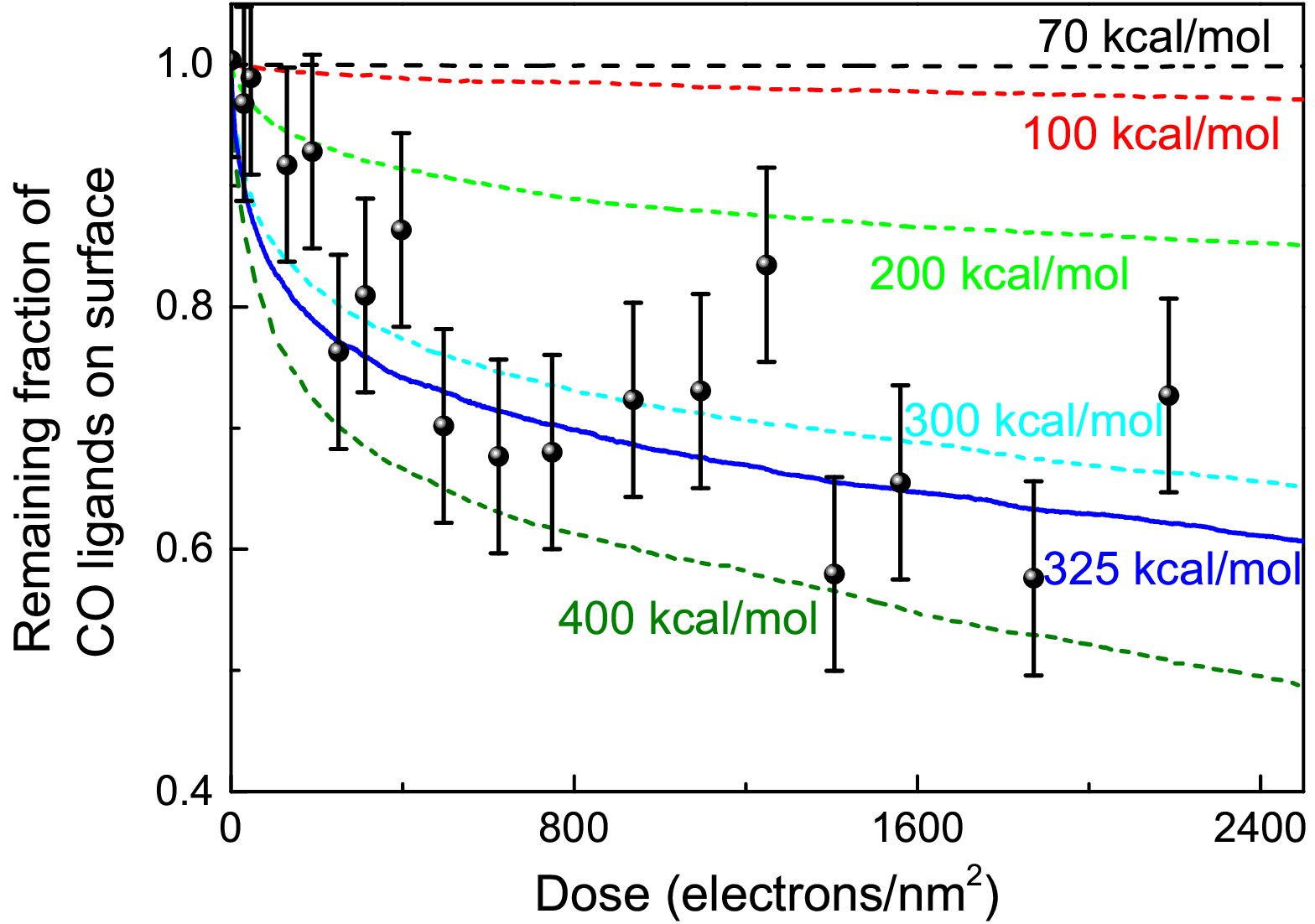}
	\caption{\label{fig3:Edep} 
		Fraction of CO ligands remaining on a gold substrate covered with W(CO)$_6$ molecules as a function of the PE dose delivered by a uniform 500 eV beam. Symbols represent experimental data \cite{Rosenberg2013}, while lines are simulation results using the indicated average deposited energies per collision.
	}
\end{figure}

The measured fraction of CO ligands remaining on the Au substrate
as a function of the PE dose 
is shown by symbols in Fig. \ref{fig3:Edep}. IDMD simulations have been performed 
for an spatially uniform 
beam, employing the 
experimentally determined decomposition cross section $\sigma_{\rm decomp}(T_0)$ for W(CO)$_6$ molecules on gold \cite{Rosenberg2013}. Note that, in this particular case, there is no need to rely on MC simulation results, 
since the fragmentation rate $P = J_0 \, \sigma_{\rm decomp}(T_0)$
is constant all over the simulated surface. The value $E_{\rm dep}$ has been scanned 
in the range 70--400 kcal/mol, consistent with gas-phase findings \cite{Cooks1990,Wnorowski2012}. As can be seen from the simulated curves, 
a value of 300--325 kcal/mol ($\sim$13--14 eV) reproduces fairly well the main features of the experimental data, this value reasonably agreeing with 
the average deposited energies estimated from gas-phase mass spectrometry \cite{Cooks1990,Wnorowski2012}.
This comparison further supports our choice for $E_{\rm dep}$.

\subsection{Models for the chemistry of reactive sites within the growing nanostructures}

Once molecules are fragmented and chemically reactive sites start to appear in the system, atoms with dangling bonds will start to form new bonds as they encounter other atoms with unsaturated valencies, following the rules of the reactive force field \cite{Sushko2015}. In a previous work \cite{Sushko2016}, two models for the formation of new chemical bonds were introduced: model A  only accounts for the chemistry between newly added precursors and the growing deposit, without restructuring of the dangling bonds within the latter (i.e., the search for reactive neighbours is done only among the atoms located beyond the molecular structure to which a chosen atom belongs). Model B allows for the formation of new bonds within the deposit itself (i.e., the search is performed over all atoms with dangling bonds in the simulation box, including the molecular structure to which a chosen atom belongs). As can be seen in the main text, consideration of these two different chemistry models A and B can lead to slightly different deposit compositions. However, the variations observed in simulations are similar to, or even lower than, the deviations observed in experimental measurements \cite{Porrati2009}, so the accuracy of the present simulations is comparable to that of available experimental techniques.


\bibliography{library}
\bibliographystyle{naturemag}
